\documentclass[superscriptaddress,preprintnumbers,amsmath,amssymb,twocolumn,floats]{revtex4-1}
\usepackage{txfonts}
\usepackage{amssymb}
\usepackage{graphicx} 
\usepackage{sidecap}
\usepackage{CJK}
\usepackage{blindtext}
\usepackage{url}
\usepackage{stfloats}
\usepackage{multirow}
\usepackage[table,xcdraw]{xcolor}
\usepackage{epstopdf} 
\usepackage{color}
\usepackage{mathtools}
\usepackage{amsmath,amssymb}
\usepackage{mathrsfs}
\usepackage{natbib}
\usepackage{bbm}

\usepackage{amsmath}
\numberwithin{equation}{section}
\begin{document}
\bibliographystyle{naturemag}
\title{Direct observation of topological surface states in the layered kagome lattice with broken time-reversal symmetry}

\author{Zhicheng Jiang}
\thanks{Equal contributions}
\affiliation{State Key Laboratory of Functional Materials for Informatics, Shanghai Institute of Microsystem and Information Technology, Chinese Academy of Sciences, Shanghai 200050, China}

\author{Tongrui Li}
\thanks{Equal contributions}
\affiliation{National Synchrotron Radiation Laboratory, University of Science and Technology of China, Hefei, Anhui 230029, China}

\author{Jian Yuan}
\thanks{Equal contributions}
\affiliation{School of Physical Science and Technology, ShanghaiTech University, Shanghai 201210, China}

\author{Zhengtai Liu}
\thanks{Equal contributions}
\thanks{ztliu@mail.sim.ac.cn}
\affiliation{Shanghai Synchrotron Radiation Facility, Shanghai Advanced Research Institute, Chinese Academy of Sciences, Shanghai 201210, China}

\author{Zhipeng Cao}
\affiliation{National Laboratory of Solid State Microstructures, School of Physics, Collaborative
Innovation Center of Advanced Microstructures, Nanjing University, Nanjing 210093, China}

\author{Soohyun Cho}
\affiliation{National Synchrotron Radiation Laboratory, University of Science and Technology of China, Hefei, Anhui 230029, China}

\author{Mingfang Shu}
\affiliation{Key Laboratory of Artificial Structures and Quantum Control (Ministry of Education), Shenyang National Laboratory for Materials Science, School of Physics and Astronomy, Shanghai Jiao Tong University, Shanghai 200240, China}
\affiliation{National Laboratory of Solid State Microstructures, School of Physics, Collaborative Innovation Center of Advanced Microstructures, Nanjing University, Nanjing 210093, China}

\author{Yichen Yang}
\affiliation{State Key Laboratory of Functional Materials for Informatics, Shanghai Institute of Microsystem and Information Technology, Chinese Academy of Sciences, Shanghai 200050, China}

\author{Jianyang Ding}
\affiliation{State Key Laboratory of Functional Materials for Informatics, Shanghai Institute of Microsystem and Information Technology, Chinese Academy of Sciences, Shanghai 200050, China}

\author{Zhikai Li}
\affiliation{School of Physical Science and Technology, ShanghaiTech University, Shanghai 201210, China}

\author{Jiayu Liu}
\affiliation{State Key Laboratory of Functional Materials for Informatics, Shanghai Institute of Microsystem and Information Technology, Chinese Academy of Sciences, Shanghai 200050, China}

\author{Zhonghao Liu}
\affiliation{State Key Laboratory of Functional Materials for Informatics, Shanghai Institute of Microsystem and Information Technology, Chinese Academy of Sciences, Shanghai 200050, China}

\author{Jishan Liu}
\affiliation{Shanghai Synchrotron Radiation Facility, Shanghai Advanced Research Institute, Chinese Academy of Sciences, Shanghai 201210, China}

\author{Jie Ma}
\affiliation{Key Laboratory of Artificial Structures and Quantum Control (Ministry of Education), Shenyang National Laboratory for Materials Science, School of Physics and Astronomy, Shanghai Jiao Tong University, Shanghai 200240, China}
\affiliation{National Laboratory of Solid State Microstructures, School of Physics, Collaborative
Innovation Center of Advanced Microstructures, Nanjing University, Nanjing 210093, China}

\author{Zhe Sun}
\affiliation{National Synchrotron Radiation Laboratory, University of Science and Technology of China, Hefei, Anhui 230029, China}

\author{Yanfeng Guo}
\email{guoyf@shanghaitech.edu.cn}
\affiliation{School of Physical Science and Technology, ShanghaiTech University, Shanghai 201210, China}
\affiliation{ShanghaiTech Laboratory for Topological Physics, ShanghaiTech University, Shanghai 201210, China}

\author{Dawei Shen}
\email{dwshen@ustc.edu.cn}
\affiliation{National Synchrotron Radiation Laboratory, University of Science and Technology of China, Hefei, Anhui 230029, China}

\begin{abstract}
Magnetic topological quantum materials display a diverse range of fascinating physical properties which arise from their intrinsic magnetism and the breaking of time-reversal symmetry. However, so far, few examples of intrinsic magnetic topological materials have been confirmed experimentally, which significantly hinder our comprehensive understanding of the abundant physical properties in this system. The kagome lattices, which host diversity of electronic structure signatures such as Dirac nodes, flat bands, and saddle points, provide an alternative and promising platform for in-depth investigations into correlations and band topology. In this article, drawing inspiration from the stacking configuration of MnBi$_2$Te$_4$, we conceive and then synthesize a high-quality single crystal EuTi$_3$Bi$_4$, which is a unique natural heterostructure consisting of both topological kagome layers and magnetic interlayers. We investigate the electronic structure of EuTi$_3$Bi$_4$ and uncover distinct features of anisotropic multiple Van Hove singularitie (VHS) that might prevent Fermi surface nesting, leading to the absence of a charge density wave (CDW). In addition, we identify the topological nontrivial surface states that serve as connections between different saddle bands in the vicinity of the Fermi level. Combined with calculations, we establish that, the effective time-reversal symmetry S=$\theta$$\tau_{1/2}$ play a crucial role in the antiferromagnetic ground state of EuTi$_3$Bi$_4$, which ensures the stability of the topological surface states and gives rise to their intriguing topological nature. Therefore, EuTi$_3$Bi$_4$ offers the rare opportunity to investigate correlated topological states in magnetic kagome materials.

\end{abstract}

\maketitle
\clearpage
%---------------------------introduction-------------------------
%\section{Introduction}

In solids, the reservation of symmetries defines kinds of topological invariants, which classify various topological quantum materials. Among them, the time-reversal symmetry (TRS) is most fundamental, as exemplified by two-dimensional quantum spin hall insulators and three-dimensional topological insulators (TIs), in which the preservation of TRS protects the exotic helical topological edge and surface states, respectively~\cite{Bernevig2006Quantum,Fu2007Topological}. Nevertheless, the breaking of TRS in topological quantum materials would usually give rise to more unexpected exotic topological states, including magnetic Wely semimetals~\cite{wan2011topological,wang2016time,liu2019magnetic,kuroda2017evidence}, quantum anomalous Hall insulators~\cite{yu2010quantized,chang2013experimental}, and axion insulators~\cite{li2010dynamical,xiao2018realization,mogi2017magnetic}, promoting remarkable progresses in both fundamental researches and spintronics related applications~\cite{vsmejkal2018topological}. In this regard, it is always highly desired to achieve more magnetic topological states with broken TRS.

However, given that early tremendous efforts to introduce magnetism into TIs with fine-tuned magnetic element doping were eventually proven to inevitably lead to inhomogeneous clustering of dopants~\cite{sessi2016dual,lee2015imaging}, this strategy naturally gave way to searching for intrinsic magnetic topological materials without external dosage of magnetic elements~\cite{gong2019experimental,li2019intrinsic,hirahara2017large}, which is promising to provide a much cleaner platform to study magnetic topological states. 
While, despite many theoretical proposals, very few examples of intrinsic magnetic topological insulators$\backslash$semimetals have been confirmed experimentally~\cite{li2019dirac,li2020giant}. To date, the best-known example MnBi$_2$Te$_4$ is a naturally occurring heterostrucure of magnetic MnTe and topological insulating Bi$_2$Te$_3$ layers~\cite{otrokov2017highly,gong2019experimental}. The combination of TRS and primitive cell translation symmetry protects the helical topological surface Dirac states parallel to the antiferromagnetic order even in the presence of broken TRS~\cite{otrokov2019prediction,chen2019intrinsic,zhang2019topological,li2019intrinsic}.

%such as MnBi$_2$Te$_4$ family and EuCd$_2$As$_4$, which have been confirmed. 
Kagome lattices, which host diversity of electronic structure signatures such as Dirac points, flat bands and saddle points, provide the playground for in-depth investigations into band topology and correlations~\cite{mazin2014theoretical,chisnell2015topological}. Specifically, their band characteristic structure, including flat and Dirac-like bands in the vicinity of Fermi level ($E_F$), has been identified to feature $Z_{2}$ topological invariant~\cite{ye2018massive,kang2020dirac}. Besides, a diversity of correlated electronic phases have been discovered in kagome materials, such as the magnetic order~\cite{nakatsuji2015large,liu2018giant,yin2020quantum,teng2022discovery}, nematicity~\cite{xiang2021twofold,nie2022charge,xu2022three} and unconventional superconductivity~\cite{ortiz2021superconductivity,yin2021superconductivity,ortiz2020cvs}. In this sense, kagome-latticed compounds would naturally server as an alternative and appropriate incubator for intriguing magnetic topological states.

In this article, drawing inspiration from the stacking configuration of MnBi$_2$Te$_4$, we conceive and then successfully synthesize a high-quality single crystal of EuTi$_3$Bi$_4$, a member of the AM$_3$X$_4$ family with typical kagome lattices~\cite{ovchinnikov2019bismuth,ortiz2023YVS}. Transport measurements unveil the out-of-plane antiferromagnetic (AFM) ground state therein, confirming its broken TRS at low temperatures. By combining high-resolution angle-resolved photoemission spectroscopy (ARPES) with first-principles calculations, we extensively investigate its low-lying electronic structure and reveal multiple anisotropic Van Hove singularities (VHSs) in the pseudo-hexagonal Brillouin zone (BZ). Additionally, we not only identify spectroscopic fingerprints of topologically nontrivial surface states connecting different VHSs but also provide evidence for their persistence across magnetic phase transitions. We discover that, despite the broken TRS in the ground state of EuTi$_3$Bi$_4$, the effective TRS introduced by the unique crystal structure preserves these topological surface states and then lead to a nontrivial $z_{4p}$ topological classification. Our work reveals the anisotropic VHSs and robust topological surface states in EuTi$_3$Bi$_4$, thereby laying the groundwork for further understanding of unique correlated topological physics in kagome lattices.

\begin{figure*}[t!]
\centerline{\includegraphics[width=2\columnwidth,angle=0]{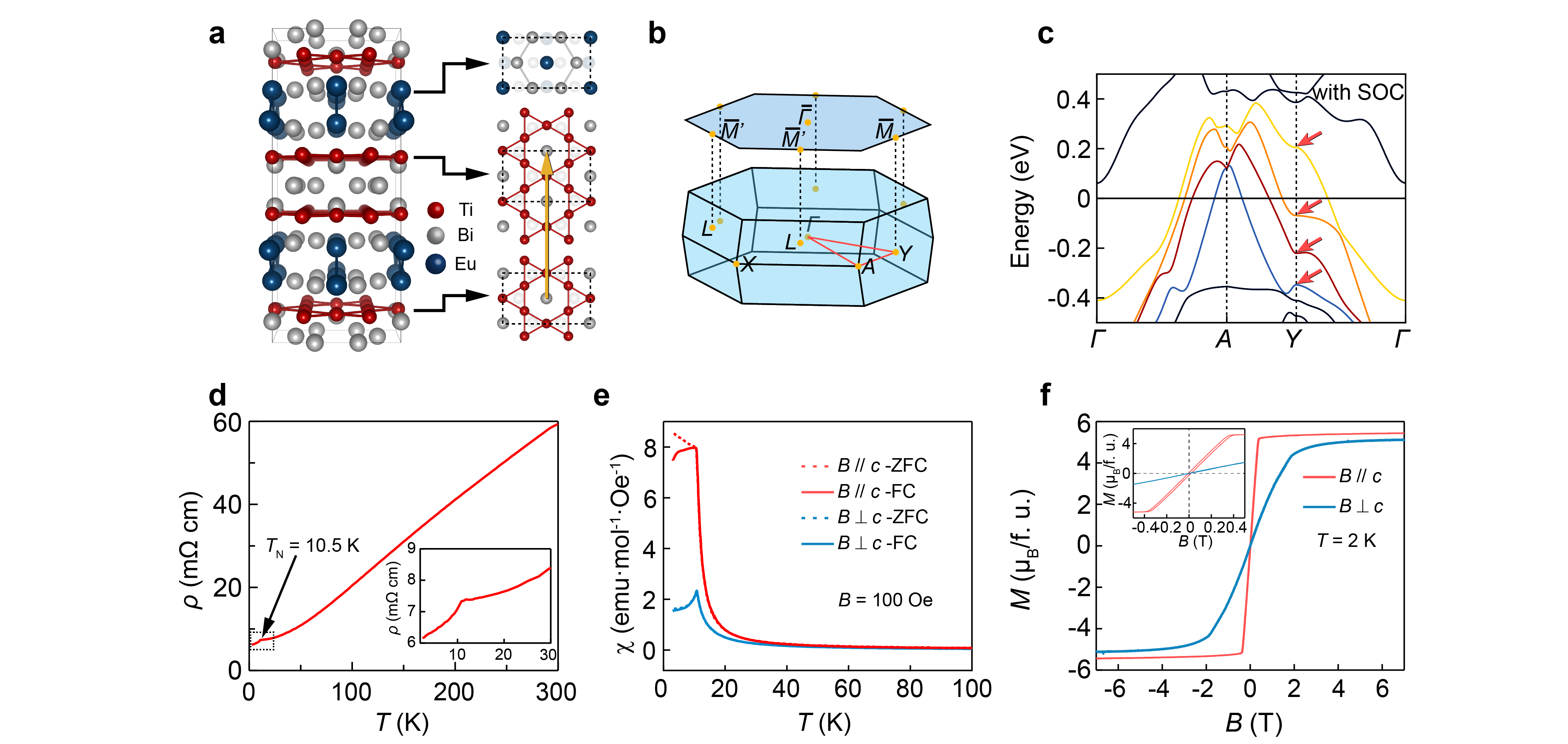}}
\caption{
\textbf{Structural and transport properties of the multi-layer titanium-based kagome compounds.}
\textbf{a} Crystal structure with uint cell of EuTi$_3$Bi$_4$;
\textbf{b} Three-dimensional Brillouin zone with its 2D projected Brillouin zone;
\textbf{c} DFT calculated band structure along the $\Gamma-A-Y-\Gamma$ direction of EuTi$_3$Bi$_4$ with spin-orbital coupling;
\textbf{d} The electronic resistivity of EuTi$_3$Bi$_4$, the inset zooms in around T$_N$;
\textbf{e} Zero-field-cooled (ZFC) and field-cooled (FC) magnetic susceptibility versus temperature with $B$ = 0.01 T;
\textbf{f} Field-dependent magnetization curves for the two directions, measured at 2 K.
}
\label{Fig 1}
\end{figure*}

\begin{figure*}[t!]
\centerline{\includegraphics[width=2\columnwidth,angle=0]{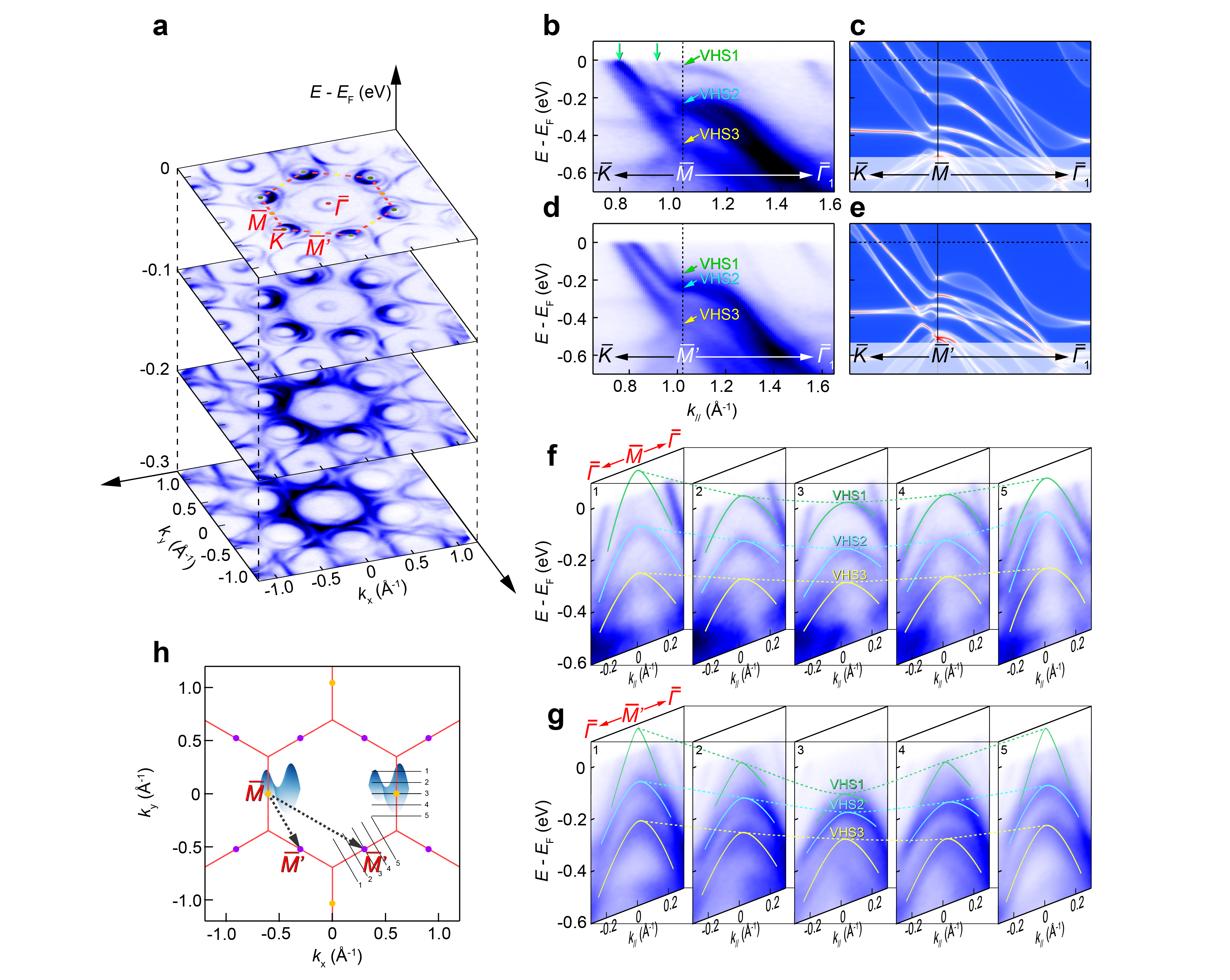}}
\caption{
\textbf{Anisotropic Van Hove Signatures (vHS) in EuTi$_3$Bi$_4$.}
\textbf{a} The stacking constant energy maps of EuTi$_3$Bi$_4$ from $E_F$ to $E_F$ - 0.3 eV;
\textbf{b} ARPES intensity plot taken along $\overline{K}$-$\overline{M}$-$\overline{\Gamma}$ direction;
\textbf{c} Calculated bulk band structure along the same direction in b;
\textbf{d} ARPES intensity plot taken along $\overline{K}$-$\overline{M}^{\prime}$-$\overline{\Gamma}$ direction;
\textbf{e} Calculated bulk band structure along the same direction in d;
\textbf{f-g} Experimentally identified VHs dispersions in EuTi$_3$Bi$_4$ by ARPES. The panels 1-5 in f (g) plot the ARPES spectra measured perpendicular to the $\overline{K}$-$\overline{M}$-$\overline{\Gamma}$ ($\overline{K}$-$\overline{M}^{\prime}$-$\overline{\Gamma}$) direction, with the cut 3 crossing the $\overline{M}$ ($\overline{M}^{\prime}$);
\textbf{h} Schematic of the in-plane Brillouin zone of EuTi$_3$Bi$_4$ (red hexagonal frames). The non-equivalent high-symmetried momentum on the center of Brillouin zone edges are marked with $M$ and $M^{\prime}$. The cut lines around $M$ and $M^{\prime}$ in f-g are marked with 1-5.
}
\label{Fig 2}
\end{figure*}

\begin{figure*}[t!]
\centerline{\includegraphics[width=2\columnwidth,angle=0]{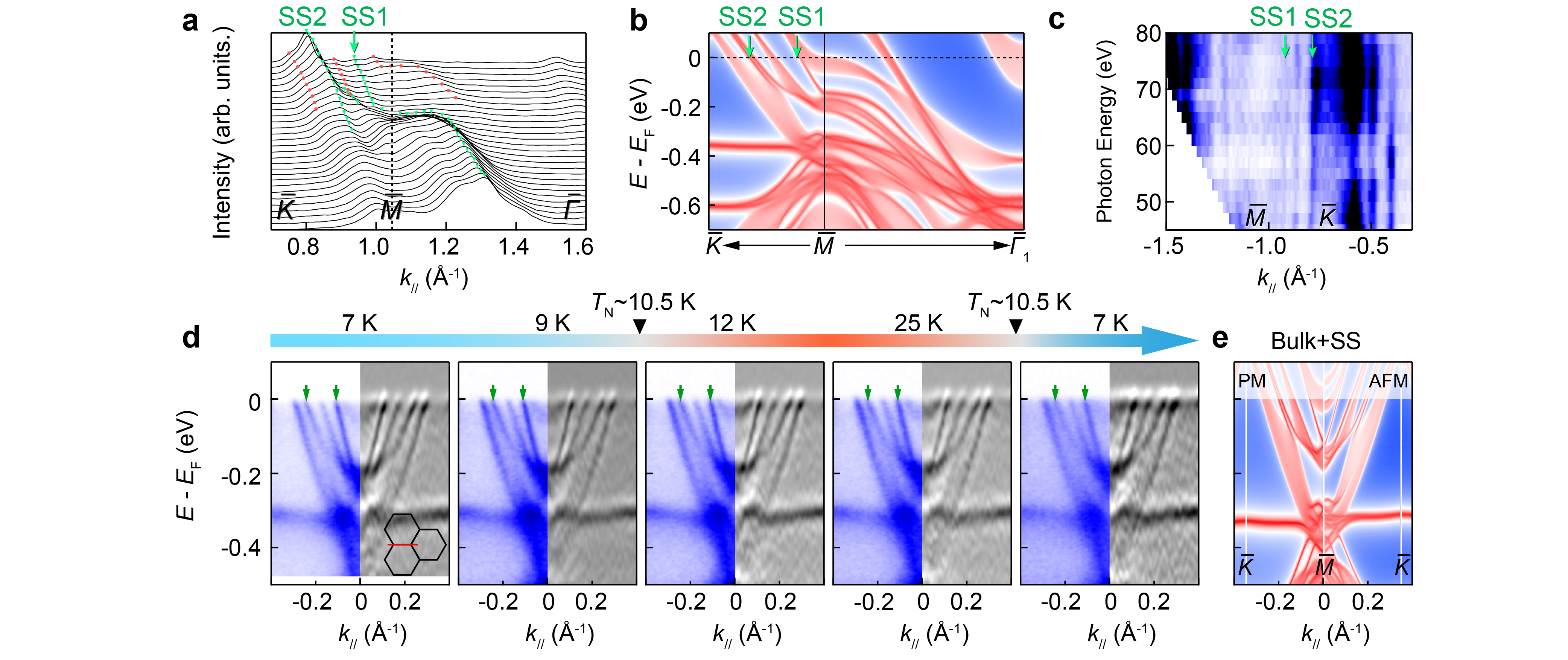}}
\caption{
\textbf{Multiple surface states in EuTi$_3$Bi$_4$.}
\textbf{a} Momentum distribution curves (MDCs) around the Van Hove signatures along the $\overline{K}$-$\overline{M}$-$\overline{\Gamma}$ direction, with the red dots mark the bulk bands and the green triangles mark the surface states (SS1 and SS2);
\textbf{b} Spectrum calculation of the bulk and surface state along the $\overline{K}$-$\overline{M}$-$\overline{\Gamma}$ direction;
\textbf{c} Photon energy dependent measurements of the ARPES spectra at $E_F$ along $\overline{K}$-$\overline{M}$-$\overline{\Gamma}$ direction in a range of 45 to 80 eV, with surface state SS1 and SS2 marked by green arrows;
\textbf{d} Side by side comparison of the temperature evolution of ARPES intensity plots along the $\overline{K}$-$\overline{M}$-$\overline{K}$ direction (left half panels) and their corresponding curvature derivative spectra (right half panels). The temperature sequence is set to 7 K, 9 K, 12 K, 25 K and return back to 7 K;
\textbf{e} Spectrum calculation with the surface states. Left half panel in e is paramagnetic (PM) state, right half panel in e is antiferromagnetic (AFM) state.
}
\label{Fig 3}
\end{figure*}

\begin{figure*}[t!]
\centerline{\includegraphics[width=2\columnwidth,angle=0]{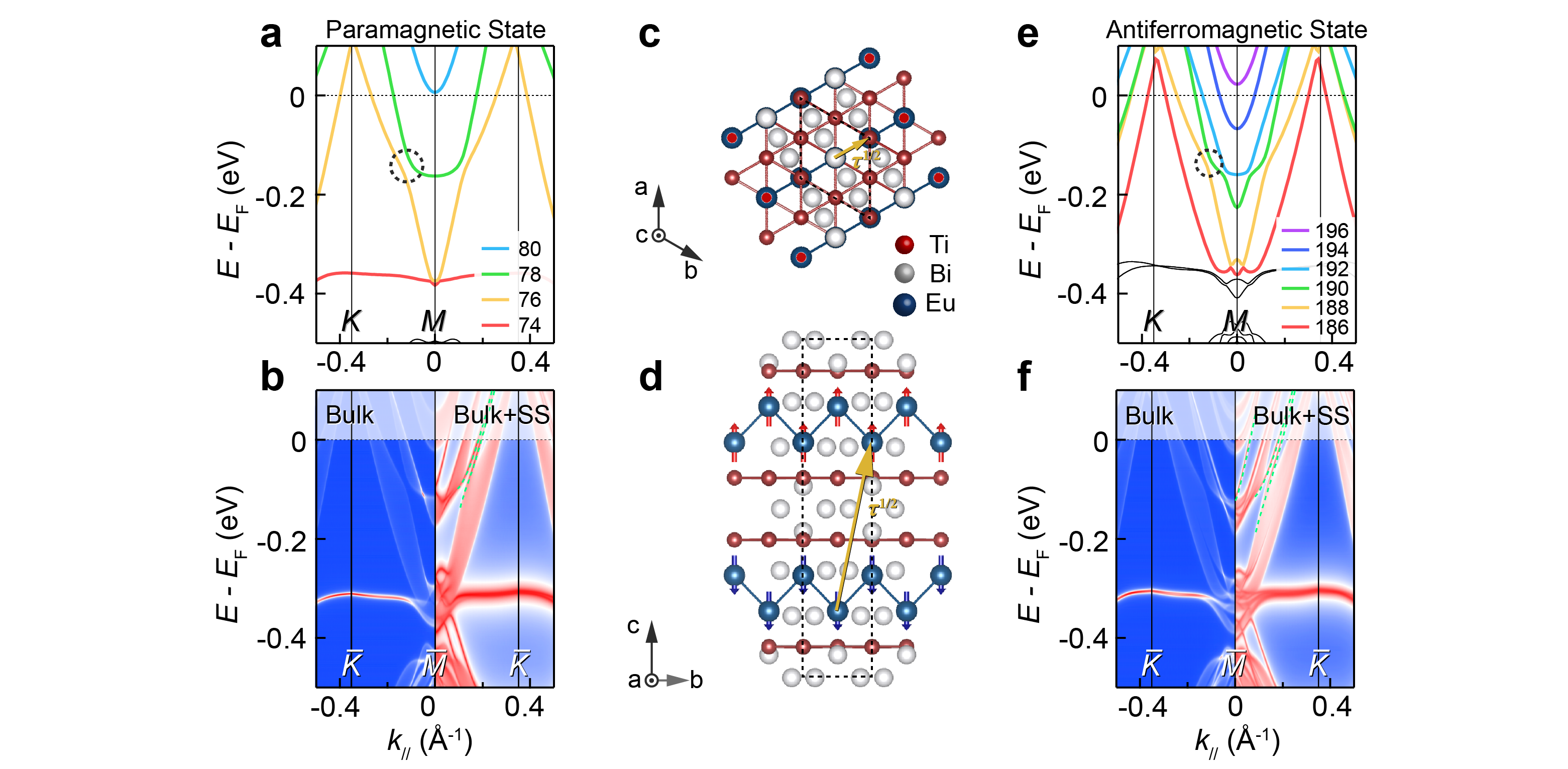}}
\caption{
\textbf{Topological properties of EuTi$_3$Bi$_4$.}
\textbf{a} Bulk band calculation in paramagnetic state and different occupied band index are marked with different colors;
\textbf{b} Spectrum calculation in paramagnetic state, left panel is bulk projected calculation, right panel is surface states calculation;
\textbf{c} Top view of the translation half-lattice vector of the antiferromagnetic state;
\textbf{d} frontal view of translated half-lattice vector  of the antiferromagnetic state;
\textbf{e} Bulk band calculation in antiferromagnetic state and different occupied band index are marked with different colors;
\textbf{f} Spectrum calculation in antiferromagnetic state, left panel is bulk projected calculation, right panel is surface states calculation.
}
\label{Fig 4}
\end{figure*}

~\\

\noindent\textbf{Results}

\noindent\textbf{Crystal and magnetic nature of EuTi$_3$Bi$_4$ single crystals.} 

EuTi$_3$Bi$_4$ crystallizes in the orthorhombic space group No.69 ($Fmmm$) with lattice constants of a = 5.9957(15) $\AA$, b = 10.422(3) $\AA$, and c = 25.474(8) $\AA$. In contrast to other typical kagome lattices such as AM$_3$X$_5$ ($P6/mmm$) and AM$_6$X$_6$ ($P6/mmm$), EuTi$_3$Bi$_4$ possesses four kagome layers in one unit cell. The crystal is primarily composed of [Ti$_3$Bi$_4$] layers stacking along the $c$ axis, with Eu atoms situated between these layers (Fig.~1a). In this way, Eu atoms form a zigzag sublattice that intertwines with honeycomb layers of Bi atoms and staggered layers of Ti-based kagome networks. Figure.~1b shows the three-dimensional orthorhombic BZ together with its projection along the [0001] direction. We note that the presence of the orthorhombic EuBi$_2$ layer effectively reduces the crystalline symmetry of EuTi$_3$Bi$_4$ to D$_{2h}$, which differs from the well-known D$_{6h}$ symmetry in AV$_3$Sb$_5$. In Fig.~1c, we display the result of first-principles calculations for the bulk band structure of EuTi$_3$Bi$_4$, which reveals the presence of multiple van Hove singularities (VHS) in the vicinity of Fermi level ($E_F$) at the high-symmetry point $Y$, along with flat bands at around 0.4 eV below $E_F$. These characteristics are commonly observed in kagome lattices, including AV$_3$Sb$_5$.

The temperature dependent electrical resistivity presented in Fig.~1d shows an anomaly near $\sim$ 10.5 K, which coincides well with the peak in magnetic susceptibility $\chi(T)$ curve (Fig.~1e), signifying the formation of long-range magnetic order. Magnetic anisotropy is observed below 10.5 K, seen by the clear different magnitude between $\chi_{ab}$ and $\chi_c$ ($\chi_c/\chi_{ab}$ $\sim$ 3.4 at 10.5 K). Under 100 Oe magnetic field, $\chi_{ab}$ drops rather sharply below 10.5 K, indicative of an antiferromagnetic-like order~\cite{Jiang2009metamagnetic}. Additionally, near the magnetic transition temperature, ZFC and FC magnetization curves exhibit overlapping peaks when B $\perp$ c, but diverge when B // c. This result suggests relatively weak interlayer magnetic coupling, facilitating the transition from the antiferromagnetic (AFM) state to the ferromagnetic (FM) state. Figure~1f presents the magnetization versus both in-plane and out-of-plane applied field, showing a quicker saturation along the $c$-axis, which indicates that the magnetic easy axis is along the $c$-axis. The inset of Fig.~1f demonstrates a tiny hysteresis loop, which can be accounted by magnetic configurations of either weak FM or AFM with small spin moments canting~\cite{felner2005possible,nayak2013large}. Note that the hysteresis loop center slightly shifts away from the zero point, suggesting the possible alternating AFM and FM domains in samples~\cite{Mei1956New}.

~\\
\noindent\textbf{Anisotropic multiple Van Hove singularities in EuTi$_3$Bi$_4$.} 

In Fig.~2, we present a comprehensive investigation of the low-energy electronic structure of EuTi$_3$Bi$_4$ through both ARPES and first-principles calculations. By examining a waterfall plot for a wide range of photon energies at a constant binding energy, a rather weak $k_z$ dispersion (see Supplementary Fig.~1) is observed, and thus we use the projected 2D BZ ($\overline{\Gamma}$, $\overline{K}$, $\overline{M}$, $\overline{M'}$) hereafter. The stacked plots of constant energy contours at different binding energies are compared in Fig. 2a. The sophisticated Fermi surface topology of EuTi$_3$Bi$_4$ is mainly composed of one inner circular and one larger hexagonal electron pockets around the center of BZ, and multiple triangular hole pockets centred at the corner of BZ, respectively. Despite the roughly hexagonal symmetric character of the FS stemming from the kagome lattice, upon one closer examination, we could still identify an anisotropy of state distribution near $E_F$ in the projected BZ: the outermost triangular Fermi pockets centered at $K$ nearly touch at $\overline{M}$, yet are far apart from each other at $\overline{M}^{\prime}$, as highlighted in Fig.~2a. 

Such anisotropy in the low-lying electronic structure stands out even more clearly in corresponding photoemission intensity plots taken along high-symmetry directions. As shown in Fig.~2b ($\overline{\Gamma}$-$\overline{M}$-$\overline{K}$), multiple kagome-lattice derived VHSs (labeled as VHS1, VHS2 and VHS3) were discovered to be located at different binding energies, and we note that the VHS1 bands lies in close proximity to $E_F$. In sharp contrast, along another high-symmetry direction ($\overline{\Gamma}$-$\overline{M}^{\prime}$-$\overline{K}$), although VHS2 and VHS3 are situated at similar binding energies, the VHS1 were revealed to be around 200 meV below $E_F$. Notably, the first-principles calculations along these two high-symmetry cuts remarkably reproduce these singularities and their anisotropy, as comparatively illustrated in Fig.~2c and 2e. In addition, more details on these VHSs can be identified according to their band dispersions exhibiting opposite concavities along two orthogonal directions, as shown in Fig.~2f and 2g, where three VHS bands display hole-like characteristics along $\overline{\Gamma}$-$\overline{M}$ ($\overline{M}^{\prime}$), while form electron-like bands along the perpendicular $\overline{K}$-$\overline{M}$ ($\overline{M}^{\prime}$) direction.

Note that previous studies proposed that the VHS near $E_F$ in AV$_3$Sb$_5$ serves as the primary driving force for the formation of charge ordering and superconductivity therein~\cite{Cho2021emergence,wang2022origin,hu2022rich,li2022tuning,oey2022tuning}. Moreover, recent reports on one another kagome metal ATi$_3$Bi$_5$, in which no evidence of long-range charge ordering has been discovered, provide further support to this hypothesis, as the VHSs in these materials are significantly far away from $E_F$. In the case of EuTi$_3$Bi$_4$, despite the presence of VHSs rather close to the Fermi level, electron scatterings between $\overline{M}$ and $\overline{M}^{\prime}$ points via ordinary wave vectors in kagome metals are prohibited due to the anisotropy of VHSs near $E_F$, as schematically displayed in Fig.~2h. Our finding of anisotropic VHSs might offer a plausible explanation for the absence of charge ordering in EuTi$_3$Bi$_4$. 

\noindent\textbf{Observation of robust surface states in EuTi$_3$Bi$_4$}
Intriguingly, the careful comparison of band dispersion along $\overline{K}$-$\overline{M}$ obtained from ARPES (Fig.~2b) and calculation (Fig.~2c) reveals some extra features which could not be captured by the bulk band structure calculation, as highlighted by two green arrows in Fig.~2b. These bands might be thus of the surface state nature. In Fig.~3a, we can provide the greater visibility of these bands in the corresponding momentum distribution curves (MDCs), and assign them as SS1 and SS2, respectively. Based on our photon energy–dependent ARPES measurements (Fig.~3c), both bands exhibit rather weak $k_z$ dispersion over half of the out-of-plane BZ, further confirming their two-dimensional surface-state nature. Below $E_F$, we have discovered that SS2 progressively separates into two distinct branches, which become significantly pronounced at a binding energy of 180 meV and then merge into VHS2 and VHS3, respectively. In order to gain more insights into the origin of these bands, we employ an iterative Green function approach to calculate the surface states along the $\overline{\Gamma}-\overline{M}-\overline{K}$ direction. As shown in Fig.~3b and Fig. S2, two additional surface bands can be identified to connect three VHS bands, in an excellent agreement with our experiments in the vicinity of $E_F$ (Fig.~2b). Besides, we discovered that these surface bands, in fact, present a similar anisotropy to bulk VHSs; they are absent near $\overline{M}^{\prime}$ but only appear around $\overline{M}$. Notably, at $\overline{M}$, the SS1 band gives rise to a surface-VHS at around 190 meV below $E_F$, which is reminiscent of the kagome surface state previously discovered in AV$_6$Sn$_6$ systems.

The magnetic susceptibility taken with different field orientations show anisotropic characteristics, as illustrated in Figs.~1e and 1f, which is as well consistent with recent reports~\cite{ortiz2023evolution}. This drove us to investigate the interplay between magnetic phase transitions and electronic structure of EuTi$_3$Bi$_4$. To this end, we carried out detailed temperature-dependent ARPES measurements. Surprisingly, the surface states remain stable across the magnetic transition temperature (T$_N$) when the base vacuum was kept in a rather high vacuum pressure, indicating their robustness against the breaking of TRS (Fig.~3d). However, when the base vacuum pressure declined from 6$\times$10$^{-11}$ Torr to 3$\times$10$^{-10}$ Torr and maintained for thirty minutes, we found that SS1 eventually vanished completely. Even after restoring the base pressure to its original level, we did not observe the reappearance of SS1, as depicted in Fig.~S3. This finding demonstrates that SS1 is vulnerable to surface perturbation primarily attributed to the adsorption of gas molecules on the surface under a degraded vacuum environment, which would kill the delicate surface states. In stark contrast, SS2 is still visible during this process, as highlighted in both intensity plots and corresponding curvature differential spectra (Fig.~S3). Furthermore, our calculations demonstrate that SS2 originates from band inversion induced by the spin-orbit coupling (SOC) and suggest its nontrivial topological characteristics, regardless of whether it is in a paramagnetic or magnetic ordered phase (Fig.~3e). Accordingly, it is highly likely that, despite the breaking of time-reversal symmetry in the magnetic ordered phase, the topological properties of SS2 should be protected by other symmetries. Given the intricate hyposymmetric crystal structure of EuTi$_3$Bi$_4$, we will delve into a detailed discussion of the origin of these surface states in the subsequent section.

\noindent\textbf{Topology analysis of the electronic structure in PM and AFM state.}
%%注释部分放SM%%

As this material is in the paramagnetic (PM) phase with the reservation of both time-reversal and inversion symmetries, we can use the Fu-Kane like formulae~\cite{prb7604,Song2018} to determine $z_{4}$ and $z_{2,j=1,2,3}$ indicators:
\begin{equation}
z_{4}=\sum_{k, k \in \mathrm{TRIM}} \frac{n_{k}^{-}-n_{k}^{+}}{2} \bmod 4\\
\end{equation}
\begin{equation}
z_{2, j}=\sum_{k \in \operatorname{TRIM}, k_{j}=\pi} n_{k}^{-} \bmod 2
\end{equation}
where $n_{k}^{-}$ and $n_{k}^{+}$ represent numbers of odd and even
parity Kramer pairs of occupied bands at the time-reversal-invariant momenta (TRIM), respectively. First of all, according to our calculation, the whole system has 80 valence electrons in the PM state, and the overall performance is the strong topological non-trivial property with $z_{4}=3$ (See details in Table \uppercase\expandafter{\romannumeral1} of SM). In order to study the topological properties of these surface states in detail, we first determined that SS2 is produced by a hybrid gap between the 76th and 78th bulk bands as shown in Fig.~4a and 4b (green dotted line). We found that the 76th bulk band has a non-zero strong topological indicators with $z_{4}=1$, so SS2 is topologically nontrivial in nature on the $\Gamma$ to $K$ high symmetry path, but the Dirac point near the $\Gamma$ point merge with the bulk band and they are difficult to distinguish. In addition, SS1 is produced at the point $M$ of the 78th and 80th bulk bands with $z_{4}=0$, so we could determine that it is of topologically trivial nature on the $\Gamma$ to $M$ high symmetry path(See details in SM).

In fact, we can not rule out the possibility of ferromagnetism according to the transport results, however, there isn't observable band splitting in ARPES data in the magnetism temperature region, so we prefer the possibility of weak antiferromagnetism. Next, to theoretically identify the band topology of EuTi$_3$Bi$_4$ in its magnetic ground state accurately, we have considered nine most likely magnetic configurations in the antiferromagnetic (AFM) phase of EuTi$_3$Bi$_4$ (Fig.~S4). Among these configurations, states with the out-of-plane magnetic momenta are energetically favorable to those with in-plane magnetic momenta. Finally, we extracted that the magnetic configuration 1 hosts the lowest energy under the same Hubburd U condition. Accordingly, such a configuration is then determined to be 68.520 ($C_Acca$) with a $\theta$$\tau_{1/2}$ effective time-reversal symmetry. Here, $\theta$ represents the time-reversal operator, and $\tau_{1/2}=\left\{ \frac{1}{2} \frac{1}{2} \frac{1}{2}\right\} $ denotes the half translation operator connecting spin-up and spin-down of the Eu atom, as illustrated in Fig.~4c and 4d. We note that each band is still doubly degenerate obviously.

To further analyze the topological properties of EuTi$_3$Bi$_4$ in the AFM phase, we employed the so-called magnetic topological quantum chemistry (MTQC) method through the Bilbao crystallographic server~\cite{Xu2020,Bradlyn2017}. This method revealed that the occupied bands of EuTi$_3$Bi$_4$ could not be expressed as a linear combination of elementary band representations (LCEBR). In other words, EuTi$_3$Bi$_4$ with 196 electrons is a non-trivial insulator or topological with non-trivial symmetry indicators. Furthermore, we calculated the topological indicators according to the method proposed by Chen $et~al.$~\cite{prb10523}. The magnetic space group (MSG) of the lowest ground state energy is 68.520, and it is possible to calculate the topological parity-based invariant $z_{4p}$ index, which is defined as: 

\begin{equation}
z_{2 p}^{\prime}=\frac{1}{2} z_{4 p}=\sum_{\mathrm{k} \in \mathrm{TRIM}} \frac{1}{4}\left(N_{k}^{-}-N_{k}^{+}\right) \bmod 2
\end{equation}
where, $N_{k}^{-}$ and $N_{k}^{-}$ are numbers of occupied bands with odd and even parity of at TRIM, respectively. The results of $z_{2 p}^{\prime}=1$ and $z_{4p}=2$, indicate the presence of non-trivial topology, which is consistent with the MTQC identification.

Interestingly, surface states of EuTi$_3$Bi$_4$ persist even in its AFM phase, as shown in Fig.~4e and 4f (green dotted line). To determine the topological properties of these surface states in antiferromagnetic phase, we first identify SS2 and SS1 produced at the 188th bulk band and 192nd bulk band of the antiferromagnetic, respectively . Using the parity values of time-reversal momentum at 8 TRIM, we possessed the table as shown in Table \uppercase\expandafter{\romannumeral2} in SM. This result indicates that SS2 is topologically nontrivial on the $\Gamma$ to $K$ high symmetry path, while SS1 is topologically trivial on the $\Gamma$ to $M$ high symmetry path, consistent with our experimental findings.

~\\
\noindent\textbf{Discussion}

We successfully synthesize the EuTi$_3$Bi$_4$ and conduct a comprehensive exploration of its electronic structure by exploiting the combination of ARPES and DFT calculations. We identifiy multiple VHSs derived from the kagome lattice in EuTi$_3$Bi$_4$, as well as their anisotropic characteristics originating from the crystal symmetry. Additionally, we uncover two surface-state dispersions that connect different saddle bands. Remarkably, we discover that the surface bands SS2 not only remain unaffected by magnetism but also exhibit strong robustness against surface perturbations. We demonstrate that the surface band SS2 originates from hybridization gap with the effective time-reversal symmetry, arising from the half translation operator connecting spin-up and spin-down of the Eu atom in the antiferromagnetic ground state, ensures the stability of surface states with $z_{4p}$ topological   classification, reminiscent of MnBi$_2$Te$_4$. In this context, our work  unveil new topological electronic states in the kagome material EuTi$_3$Bi$_4$, and provides a new avenue for investigating the intricate interplay between magnetism, spin-orbit coupling, and topological surface states in frustrated kagome lattices.

~\\
\noindent\textbf{Methods}

\noindent\textbf{Sample growth and characterization}

The EuTi$_3$Bi$_4$ crystals were synthesized by using bismuth as the flux. High purity elements of europium rod (99.9 $\%$, Alfa Aesar), titanium powder (99.99 $\%$, Macklin), and bismuth pellet (99.999 $\%$, Aladdin) were mixed in a molar ratio of 1:1:20 and placed into an alumina crucible, which was then sealed into a quartz tube in vacuum. The assembly was heated in a furnace up to 1000 $^{\circ}$C within 10 hours, kept at that temperature for 20 hours, and then cooled down to 800 $^{\circ}$C within 5 hours. It was subsequently slowly cooled down to 500 $^{\circ}$C at a temperature decreasing rate of 2$^{\circ}$C/h. The excess bismuth was removed at this temperature by quickly placing the assembly into a high-speed centrifuge.

\noindent\textbf{ARPES experiments}

All EuTi$_3$Bi$_4$ single crystal samples were cleaved {\it in situ} at 7\,K with a base pressure of better than 6 $\times$ 10$^{-11}$\,Torr. High-resolution ARPES measurements were performed at the 03U beamline of Shanghai Synchrotron Radiation Facility (SSRF)\cite{yang2021high} with liner horizontal polarization lights~\cite{sun2020performance}. In our measurements, light's linear horizontal polarization is parallel to the ground and the incident angle on sample is $\theta$ = 45$^{\circ}$ with respect to the sample's normal direction, along with the slit direction that is perpendicular to the ground. During the experiment, the beam size is set to 15 $\times$ 15 mm$^2$. All data were acquired with a Scienta-Omicron DA30 electron analyzer. The total energy resolution was set to 10$\sim$20\,meV depending on the photon energy applied, and the angular resolution was set to be 0.2$^{\circ}$.

\noindent\textbf{Band calculations}

The electronic structure calculations for EuTi$_3$Bi$_4$ were performed using density functional theory (DFT) within the projector augmented wave (PAW) method \cite{prb5017} as implemented in Vienna ab initio Simulation Package (VASP) \cite{prb511169}. The exchange-correlation functional was based on the generalized gradient approximation of Perdew, Burke, and Ernzerhof (PBE) \cite{prl773865}. The Brillouin zone integration was carried out on a 6×6×10 Monkhorst-Pack k mesh, and the cut-off energy was set to 450 eV. The experimentally lattice parameters were utilized, and the spin-orbit coupling (SOC) effect was included in all the calculations. 
The tight binding model based on maximally
localized Wannier functions \cite{prb5612847,prb65035109} was constructed to reproduce the spectral functions with the selection of Eu d, Ti d, and Bi p orbitals in the paramagnetic state. Surface states were calculated using iterative Green function methods \cite{mpls,mpls2} as implemented in WANNIERTOOLS \cite{wuwt}. The parity analysis was performed using the IRVSP code \cite{GAO2021}. 
%According to the cleavage direction of the experiment, we performed a matrix transformation on the canonical primitive cell in paramagnetic state, and the parity value calculation and spectral function calculation all use our converted primitive cell model.%

%Furthermore, SS1 is produced at the M point of the 78th and 80th bulk bands with $z_{4}=0$ but there is a non-zero value of weak topological indicator (See details in SM Table1), it implies that there may be topological non-trivial surface states on some high symmetry paths of the two-dimensional Brillouin zone. To determine the possibility, we calculated the time-reversed polarization (TPR)\cite{prb7604} values  at $\overline \Gamma$ (0, 0) and $\overline M$ (0, 0.5) points in two-dimensional Brillouin zone. We found that all TPR values were equal to 1 and their signs remained unchanged. However, considering that the $\overline \Gamma$ - $\overline M$ path is invariant under the mirror symmetry operation. To rule out mirror-protected nontrivial phase, finally, we calculate the hybrid wannier charge centers (HWCCs)\cite{prb115202,prb035108} of the corresponding mirror in the bulk Brillouin zone. Through counting the winding number of the sum of HWCCs as shown in Fig.~S5 red lines, we obtain the mirror Chern number to be zero. Therefore, we conclude that SS1 is topologically trivial.%

In the AFM calculations, the 4f electrons of Eu were treated as valence electron. The Brillouin zone integration was performed on an 8×8×3 Monkhorst-Pack k mesh, the cut-off energy was set to 450 eV. A Hubbard-like correction with U=6eV was applied to the 4f orbital of Eu, which resulted in a good match with the energy position of the ARPES 4f energy level. To reproduce the spectral functions in the AFM state, the tight-binding model based on maximally localized Wannier functions\cite{prb5612847,prb65035109} was utilized, which included Eu d and f, Ti d and Bi p orbitals. In order to calculate ground state energy for different magnetic configurations, a denser K mesh of 18×18×3 
 was employed, and cut-off energy was set to 380 eV.

The authors declare that the main data supporting the findings of this study are available within the paper and its Supplementary Material. Extra data are available from the corresponding authors upon request.

\noindent\textbf{Acknowledgements}

We thank Prof. Xiangang Wan, Prof. Yi Zheng, Prof. Yang Liu and Prof. Dong Qian for the useful discussions. We acknowledge the support by the National Natural Science Foundation of China (Grants Nos. U2032208, 92065201, U2032213). Y. F. G was sponsored by Double First-Class Initiative Fund of ShanghaiTech University and the open projects from State Key Laboratory of Functional Materials for Informatics (Grant No. SKL2022). J.M. was supported by the National Key Research and Development Program of China (Grant No. 2022YFA1402704). Part of this research used Beamline 03U of the Shanghai Synchrotron Radiation Facility, which is supported by ME$^2$ project under Contract No.11227902 from National Natural Science Foundation of China. The authors also thank the support from Analytical Instrumentation Center (\#SPST-AIC10112914).
.

\noindent\textbf{Author contributions}
Z.C.J. and Z.T.L. performed the ARPES experiment and analyzed the resulting data. T.R.L., Z. S. and Z.P.C. performed the theoretical calculations. J.Y. and Y.F.G. synthesized the single crystals and characterized the basic properties. J.Y, M.F.S., Z.K.L. and J.M. performed the transport measurements. Y.F.G. and D.W.S. supervised the project. S.C., Y.C.Y., J.Y.D., J.Y.L.,J.S.L. contributed to the development and maintenance of the ARPES systems, beamline and related software development. Z.C.J, T.R.L., Z.T.L, D.W.S wrote the manuscript with input from all coauthors.

\noindent\textbf{Competing interests}

The authors declare no competing interests.

\bibliographystyle{naturemag}
\bibliography{ETB}

\end{document}